# Lattice diffusion and surface segregation of B during growth of SiGe heterostructures by molecular beam epitaxy : effect of Ge concentration and biaxial stress

A. Portavoce, P.Gas

*L2MP-CNRS, Univ. Aix-Marseille 3, Campus St Jérôme, 13397 Marseille Cedex 20, France*

I. Berbezier, A. Ronda,

*CRMCN-CNRS, Campus de Luminy, Case 913, 13288 Marseille Cedex 9, France*

J.S. Christensen, B. Svensson

*Royal Institute of Technology (KTH), Department of Electronics, Electrum 229, SE-164 40*

*Kista-Stockholm, Sweden*

**Abstract**

$Si_{1-x}Ge_x/Si_{1-y}Ge_y/Si(100)$ heterostructures grown by Molecular Beam Epitaxy (MBE) were used in order to study B surface segregation during growth and B lattice diffusion. Ge concentration and stress effects were separated.

Analysis of B segregation during growth shows that: i) for layers in epitaxy on (100)Si), B segregation decreases with increasing Ge concentration, i.e. with increased compressive stress, ii) for unstressed layers, B segregation increases with Ge concentration, iii) at constant Ge concentration, B segregation increases for layers in tension and decreases for layers in compression. The contrasting behaviors observed as a function of Ge concentration in compressively stressed and unstressed layers can be explained by an increase of the equilibrium segregation driving force induced by Ge additions and an increase of near-surface diffusion in compressively stressed layers.

Analysis of lattice diffusion shows that: i) in unstressed layers, B lattice diffusion coefficient decreases with increasing Ge concentration, ii) at constant Ge concentration, the diffusion coefficient of B decreases with compressive biaxial stress and increases with tensile biaxial stress, iii) the volume of activation of B diffusion ($\Delta V = -kT\frac{d\ln D}{dP}$) is positive for biaxial stress while it is negative in the case of hydrostatic pressure. This confirms that under a biaxial stress the activation volume is reduced to the relaxation volume.

# I. INTRODUCTION

The control of dopant redistribution during growth of heterostructures is a crucial issue for the realization of ultimate devices. The main physical problem to overcome is the tendency for dopant atoms to segregate to the surface during epitaxial growth. This renders difficult the formation of ultra-shallow junctions and of locally doped nanostructures such as quantum wells, wires or dots[1]. For $Si_{1-x}Ge_x$/Si heterostructures, the situation is complicated by the fact that redistribution processes will depend on both Ge concentration and compressive stress due to the $Si_{1-x}Ge_x$/Si lattice mismatch. The intrinsic influence of these two effects is unclear since experiments are often carried out with epitaxial films on Si. In a recent analysis of Sb surface segregation during MBE growth of $Si_{1-x}Ge_x$/Si heterostructures, we have separated the effects of Ge concentration and compressive stress[2]. We showed that Sb surface segregation: i) decreases with increasing Ge concentration in unstressed layers, and ii) increases with Ge concentration in compressively stressed layers (epitaxy on (100)Si). The Ge concentration and the compressive stress have thus opposite effects. Taking into account the dependence of this redistribution process on both thermodynamics (Sb segregation energy) and kinetics (Sb diffusion close to the surface), the contrasting behaviors of stressed and unstressed films were explained by a decrease of the segregation energy due to Ge addition, combined with an increase of diffusion in stressed layers. Measurements of the Sb lattice diffusion in the same layers corroborated this last statement and showed that the diffusion of Sb increases with Ge concentration in unstressed films and increases with compressive stress[3] at constant Ge concentration.

The goal of this paper is to conduct a similar analysis allowing the discrimination of concentration and stress effects on B surface segregation and diffusion. Apart from the practical interest of analyzing B redistributions, the comparison between B and Sb is interesting for several reasons:

- the three main factors which influence equilibrium surface segregation[4], namely : a) surface energies ($\gamma^B > \gamma^{Si} > \gamma^{Ge} > \gamma^{Sb}$), b) atomic sizes ($r_B < r_{Si} < r_{Ge} < r_{Sb}$), and c) the nature of interactions (compound formation for B-Si, phase separation for Sb-Si)[5] are different for B and Sb.

- moreover, the lattice diffusion mechanisms of these two dopants are different: B diffusion is mediated by interstitials while Sb diffusion uses vacancies[6].

## II. EXPERIMENT

In order to distinguish the stress effect from that of Ge concentration, we analyzed the B distribution profiles resulting from the growth of stressed and unstressed $Si_{1-x}Ge_x$ layers ($0 \leq x \leq 0.2$). These layers were grown in a Riber MBE system with a residual pressure typically ~ $10^{-11}$ torr. Silicon was evaporated using an electron gun from a floating zone silicon crystal. Germanium and boron were evaporated from effusion Knudsen cells. Boron doped Si(100) wafers of nominal orientation (misorientation < 0.2 deg) were used as substrates. They were first cleaned and protected by an oxide layer using a standard chemical process. After introduction in the growth chamber, the wafers were annealed at 900°C to dissociate the surface oxide. A 50 nm thick Si buffer layer was then grown on the substrates at 750°C in order to achieve a reproducible initial Si surface; its quality was checked by the RHEED intensity of the (2x1) reconstruction. Two sets of structures were grown on the Si buffer, they are represented on figure 1. The first set (type 1) consists of: (a) a 50 nm thick $Si_{1-x}Ge_x$ layer, (b) a B film (1/10 of monolayer), (c) a 50 nm thick $Si_{1-x}Ge_x$ layer, (d) a 20 nm thick cap of Si grown at 200°C. Layers (a), (b) and (c) were grown at 550°C while layer (d) was grown at 200°C. In these structures the $Si_{1-x}Ge_x$ layers have all the same concentration, they are thus either fully compressively stressed (x = 0.09 or x = 0.18) or unstressed (x = 0). The goal of the Si cap is to trap boron atoms, that have segregated during the growth of the SiGe layer (c), between the interface of layers (d) and (c). The second set (type 2) consists of the same (a), (b) and (c) layers but deposited on a relaxed $Si_{1-y}Ge_y$ buffer. In these structures the $Si_{1-x}Ge_x$ layers are thus either unstressed (x = y = 0.09 or x = y = 0.18) or under tension (x = 0.09, y = 0.18). The relaxed $Si_{1-y}Ge_y$ buffers were produced using the low temperature compliant layer process as described in references 2 and 3.

The Ge concentration in the layers was checked by Rutherford Backscattering Spectrometry (RBS). The B concentration versus depth profiles were measured by secondary ion mass spectrometry (SIMS) using a Cameca IMS4F with 8 Kev $O_2^+$ primary beam.

For diffusion measurements, samples of different concentrations and stress states were annealed together (in a inert $N_2$ ambient) in order to minimize experimental uncertainties. The B lattice diffusion coefficient was then deduced from a comparison between the B distribution measured after annealing and a numeric fit of the diffusion equation using the non-annealed profile as initial distribution[3].

## III. RESULTS

### A. Boron segregation during MBE growth

Figure 2 shows the B concentration profiles obtained in three $Si_{1-x}Ge_x$ structures (x = 0, 0.09, 0.18) grown at 550°C on Si(100) and covered with a Si cap (type 1); dashed lines are used to visualize the different parts of these structures. If one excepts the three or four first points which correspond to the free surface and are thus (because of the presence of a surface oxide) meaningless, one can distinguish two peaks. The first one is located at a depth of ~ 20-30 nm, thus close to the interface between layers (c) and (d). It corresponds to the quantity of B segregated during the growth of the $Si_{1-x}Ge_x$ layer (c) at 550°C and trapped at this interface. The second one is located at a depth of ~65-70 nm, thus close to the position of the B layer. It corresponds to the B incorporated quantity ($Q_{inc}$). As these layers are covered with a Si cap, it is possible to estimate both the total quantity of B ($Q_{tot}$) and the incorporated one ($Q_{inc}$) by integration of either the total SIMS profile or the right part of the profile (the limit of integration are also given on fig. 2). However, in the structures of type 2 without Si cap, such a procedure is not possible and only the incorporated quantity can be measured. In that case, we used for $Q_{tot}$ the value obtained in the capped samples.

The coefficient of incorporation of B ($r = Q_{inc}/Q_{tot}$) obtained for the different layers (under compression, unstressed, under tension) is plotted as a function of the Ge concentration in Figure 3. One observes that:

- for $Si_{1-x}Ge_x$ layers in epitaxy on (100) Si) (i.e. under compression), the coefficient of incorporation is larger than for pure Si layers. This is due to the combined effect of Ge concentration and induced compressive stress. Incorporation is almost total in the $Si_{0.91}Ge_{0.09}$ layer,

- for unstressed layers, the coefficient of incorporation decreases when Ge concentration increases,

- at constant Ge concentration (x = 0.09), the coefficient of incorporation increases with a compressive stress and decreases with a tensile stress.

Since surface segregation follows the opposite behavior of incorporation, one can conclude that: i) for unstressed layers, the addition of Ge increases the B segregation, ii) at constant Ge concentration, tensile stress increases the B segregation and compressive stress decreases it, iii) the combined effect of Ge concentration and compressive stress (layers for epitaxy on Si(100)) is to decrease B segregation. These behaviors are summarized on Figure 4, where the segregation coefficient ($s = 1 - r$) is plotted versus Ge composition and biaxial stress. The stress is expressed as the difference ($\Delta x$) between the Ge composition (at%) of the layer of interest and the Ge composition of the layer used as epitaxial substrate.

## B. Boron lattice diffusion in $Si_{1-x}Ge_x$ layers

Figure 5 shows the B concentration profiles measured in an unstressed $Si_{0.91}Ge_{0.09}$ layer (type 2 structure) after annealing at 900°C during 1 hour. The as-grown B profile and the fit to the diffused profile are also included in the figure. The fit corresponds to a diffusion coefficient equal to $3.9 \times 10^{-15}$ cm$^2$/s. The agreement between annealed and simulated profile is good in the whole dopant concentration range. The value of the diffusion coefficient is however higher, by a factor of ~2.5, than that measured recently by Zangenberg et al[7] in an unstressed layer of similar composition. Such discrepancies which are not exceptional in the analysis of dopant diffusion coefficients in Si and SiGe[6-8] can be explained by the different experimental conditions used in the two studies. Given the scope of this study (joint analysis of B diffusion and segregation), we used a relatively low growth temperature 550°C (growth temperature is not mentioned in ref 7), a Si cap or no cap at all (while a $SiO_2/Si_2N_3$ bilayer is used in ref. 6), a higher dopant concentration and a different annealing atmosphere ($N_2$ vs air).

Figure 6 shows the evolution of the diffusion coefficient of B with Ge concentration at 900°C in unstressed (x = 0, 0.09, 0.18) and compressively stressed (x = 0.09, 0.18) $Si_{1-x}Ge_x$ layers (epitaxy on (100)Si). In both cases, the B diffusion coefficient decreases when Ge concentration increases in agreement with former studies on the influence of Ge on boron diffusion in $Si_{1-x}Ge_x$ layers[6-8]. The diminution observed for layers in epitaxy on (100) Si is larger than that observed for unstressed layers, showing that both Ge concentration and induced compressive stress lead to a decrease of the B diffusion coefficient. In our case, the influence of these two factors on the observed decrease is comparable, while former studies by Kuo et al[9] and Zangenberg et al[7] on stressed and unstressed $Si_{1-x}Ge_x$ layers have respectively attributed this decrease mainly to concentration[9] or stress[7].

The influence of stress is confirmed in Figure 7 where the diffusion coefficient of B measured in $Si_{0.91}Ge_{0.09}$ is plotted as a function of stress from tension to compression. The stress state is quantified through the biaxial pressure:

$$P^b = -2\mu \frac{\nu+1}{\nu-1} \frac{a_{film} - a_{sub}}{a_{sub}} \qquad (1)$$

In this expression, valid for a film in epitaxy on a substrate, $a_{film}$, $\mu$ and $\nu$ are respectively, the lattice parameter, the shear modulus and the Poisson ratio of the film. $a_{sub}$ is the lattice parameter of the substrate. This equation gives positive pressures for biaxial compression and negative pressures for biaxial tension.

It appears from Figure 7 that the diffusion coefficient of B decreases with a biaxial compression and increases with a biaxial tension, in agreement with the work of Zangenberg et al[7]. The values of $D_B$ for the different layers are reported in Table I and presented next to the segregation coefficient on Figure 4.

## IV. DISCUSSION

The distinction between the influence of Ge concentration and epitaxial stress allows a better analysis of B surface segregation and diffusion in $Si_{1-x}Ge_x$ layers.

### A. Boron surface segregation during growth

Figure 4 shows that B segregation during the growth of $Si_{1-x}Ge_x$ layers:

i) increases with increasing Ge concentration in unstressed layers, ii) decreases with compressive stress and increases with tensile stress at constant Ge concentration, iii) decreases with increasing Ge concentration in layers for epitaxy on Si(100) (combined effect of Ge concentration and induced compressive stress).

It is important to mention that:

- the variations observed for B are opposite to that observed for Sb, e.g., Ge addition to layers for epitaxy on Si(100), decreases B segregation while it increases Sb segregation[2],

- the contrasting behavior observed as a function of Ge concentration, for B segregation in layers in epitaxy on (100)Si and in unstressed layers was also observed for Sb segregation[2].

In order to understand these features, the same analysis as the one previously used for Sb segregation can be used, namely:

- we consider the factors which affect the flux of B which segregates at the surface during the growth of $Si_{1-x}Ge_x$ layers and their dependence on Ge concentration and stress. This flux is proportional to the product of the diffusion coefficient close to the surface and the thermodynamic driving force for segregation,

- we make an analogy between the variations of the diffusion coefficient in the vicinity of the surface and that observed in the volume.

Thus, if one takes into account only the variations of the diffusion coefficient, the B segregation during growth should decrease with increasing Ge concentration, both in unstressed and compressed layers. This agrees with the results obtained in compressed layers but not in unstressed layers. In that latter case, the B segregation increases with Ge addition, which implies that the driving force for B segregation should increase with Ge addition. The main driving force for surface segregation is the decrease of surface energy[4]. For B segregation in Si (or Ge) this factor is not favorable. The surface energy of B is larger than that of Si (and Ge). Steric effects do not either favor a strong B segregation: B atoms have a smaller size than Si and Ge atoms[4]. The only driving force in favor of B surface segregation is its low solubility[10]. This solubility is lower in Ge than in Si[5], moreover one has to notice that although the Si/B phase diagram evidences compound formation, the Ge/B phase diagram shows perfect demixion. This would agree with an increase of the driving force for B segregation with Ge addition to Si.

The variations observed in unstressed layers lead thus to the conclusion that the increase of the segregation driving force has a more effective influence on B segregation than the decrease of the B diffusion coefficient. For compressed layers, the decrease of the B diffusion coefficient is larger. In that case, either the increase of the B segregation driving force is not sufficient to compensate this larger decrease of the diffusion coefficient or the addition of compressive stress induces also a decrease of the B segregation driving force. Given the small size of B, this second effect appears less probable.

It is important to notice that the conclusion drawn from the analysis of Sb segregation appears to be also valid for B. Namely, that for unstressed layers, the modification of the segregation driving force has a dominant influence while for compressively stressed layers the modification of the diffusion coefficient due to stress compensates this effect. Because these factors (segregation driving force and diffusion coefficient) are opposite for Sb and B, the modifications of surface segregation during growth are also opposite.

## B. Boron lattice diffusion

In unstressed layers, the diffusion coefficient of B decreases with increasing Ge. Since the B diffusion mechanism in $Si_{1-x}Ge_x$ was shown to be mediated via interstitials (when $x < 0.2$[8, 11-14]), this decrease indicates a diminution of the efficiency of this mechanism. This may either be due to a decrease in the interstitial concentration and mobility[6,7] or/and to a binding between Ge and B[9].

At constant Ge concentration, the B diffusion coefficient decreases with a biaxial compressive stress and increases with a biaxial tensile stress. These variations can be used to analyze quantitatively the effect of a biaxial pressure on B diffusion through the activation volume.

Providing diffusion takes place via a single diffusion mechanism, the activation volume is given by[15-16]:

$$\Delta V = -kT \left( \frac{d\ln D}{dP} \right) \qquad (2)$$

Where k is the Boltzmann constant, T the temperature and P the pressure.

The variations reported in Figure 7 lead to a positive activation volume: $\Delta V \sim 1.5\ \Omega$ (with $\Omega$ the lattice site volume). Let us recall that under hydrostatic pressure, the activation volume for B diffusion in Si is negative ($\Delta V \sim -0.2\ \Omega$ at T = 810°C)[16] which is in agreement with a diffusion mechanism via interstitial.

The situation is thus similar to that found for Sb: the activation volume under biaxial stress and hydrostatic pressure have opposite signs. This can be explained considering the theoretical model proposed by Aziz[18-20]. Considering this model, the contribution of the activation volume in the expression of the work performed by the system under hydrostatic pressure is written:

$$\Delta V^h = \pm \Omega + V^r + V^m \qquad (3)$$

while under biaxial pressure it is written:

$$\Delta V^b = \frac{2}{3} V^r + V^m - V^m_{//} \qquad (4)$$

In these expressions, the + sign is for vacancy formation and the − sign for interstitial formation, $V^r$ is the relaxation volume, $V^m$ is the trace of the migration volume tensor, $V^m_{//}$ ($V^m_{\perp}$) its component in the direction parallel (resp. perpendicular) to the direction of diffusion.

If one assumes that the volume of the defect is constant and that each direction has a constant elasticity, then $V^m_{\perp} = 0$ and $V^m_{//} = 0$, leading to $\Delta V^m = 0$. This is equivalent to neglecting the migration part of the activation volume. Within this approximation, and assuming that for an interstitial:

$$V^r \geq 0 \text{ and } |V^r| \leq \Omega \qquad (5)$$

the activation volume is found negative for an hydrostatic pressure:

$$\Delta V^h \approx -\Omega + V^r \qquad (6)$$

and positive for a biaxial pressure:

$$\Delta V^b \approx \frac{2}{3} V^r \qquad (7)$$

since under biaxial pressure, the activation volume contributing to the work contains only the relaxation part. It is important to stress that the parallel analysis of two dopants (Sb and B) which diffuse by different mechanisms leads to this similar conclusion and to the fact that a hydrostatic pressure and a biaxial pressure have opposite effects on diffusion.

## V. CONCLUSIONS

We used $Si_{1-x}Ge_x/Si_{1-y}Ge_y/Si(100)$ heterostructures prepared via MBE to study the B lattice diffusion in $Si_{1-x}Ge_x$ layers (at 900°C) and the B surface segregation during $Si_{1-x}Ge_x$ growth (at 550°C). The influence of Ge concentration and biaxial stress were separated.

For B diffusion, we show that:

i) in unstressed layers, the B lattice diffusion coefficient decreases when Ge concentration increases ($0 \leq x \leq 0.2$),

ii) at constant Ge concentration, the B diffusion coefficient decreases with compressive biaxial stress and increases with tensile biaxial stress,

iii) in layers for epitaxy on Si(100), the B lattice diffusion coefficient decreases when Ge concentration increases. This decrease is thus due to the cooperative effect of Ge concentration and induced compressive biaxial stress.

The volume of activation of B diffusion deduced from the variation of the B diffusion coefficient versus biaxial stress is positive (about $1.5 \Omega$ at 900°C in a $Si_{0.91}Ge_{0.09}$ layer) while it is negative under hydrostatic pressure. This is in agreement with an activation volume reduced to the relaxation volume under biaxial stress. Thus, for the same diffusion mechanism, biaxial compression and hydrostatic pressure have opposite effects. This is observed for B as well as for Sb diffusion[3].

For B surface segregation during MBE growth, we show that:

i) in compressively stressed layers (epitaxy on (100)Si), B segregation decreases when Ge concentration increases,

ii) in unstressed layers, B segregation increases with Ge concentration,

iii) at constant Ge concentration, B segregation increases for layers under tension and decreases for layers under compression.

We analyze these variations on the basis of a kinetic limitation of the B segregation. The contrasting behaviors observed as a function of Ge concentration in stressed and unstressed layers can thus be explained by a decrease of diffusion in stressed layers combined with an increase of the segregation driving force due to the addition of Ge. For unstressed layers, the modification of the segregation driving force has a dominant influence, while for compressively stressed layers, the additional modification of the diffusion coefficient due to stress compensates this effect. Similar conclusions have been drawn from the analysis of Sb segregation during $Si_{1-x}Ge_x$ layer growth[2].

## ACKNOWLEDGMENTS

We are very grateful to F.M. d'Heurle (IBM T.J. Watson Research Center, NY) and D. Mangelinck (L2MP, Marseille) for discussion and critical reading of the manuscript. We also would like to thank A. Yu. Kuznetsov of the Royal Institute of Technology (Stockholm) for his help.

# FIGURE CAPTION

**Figure 1.** Schematics of type 1 and type 2 structures: (a) 50 nm thick $Si_{1-x}Ge_x$ layer, (b) B film (1/10 of monolayer), (c) 50 nm thick $Si_{1-x}Ge_x$ layer and (d) 20 nm thick cap of Si (only type 1).

**Figure 2.** Depth concentration profiles of B in $Si_{1-x}Ge_x$ layers grown at 550°C on a Si(100) substrate (Type 1 structures): x = 0 (squares), x = 0.09 (circles), x = 0.18 (triangles).

**Figure 3.** Coefficient of incorporation of B during the growth of $Si_{1-x}Ge_x$ layers at 550°C. Variations with Ge concentration for different biaxial stress conditions: layers in compression (full squares), unstressed layers (full circles), layer in tension (open square). The error bars take into account the uncertainty on $Q_{tot}$ for type 2 structures.

**Figure 4.** Variations of the B segregation coefficient at 550°C (s) and of the B diffusion coefficient at 900°C (D) in $Si_{1-x}Ge_x$ layers versus stress and Ge concentration. The stress ($\Delta x = x_l - x_s$) is expressed as the difference between the Ge composition (at%) of the layer of interest and the Ge composition of the layer used as epitaxial substrate. The segregation coefficient "s" is deduced from the coefficient of incorporation "r" (s = 1 – r) presented in figure 3. The diffusion coefficient "D" is expressed using $10^{-15}$ $cm^2/s$ as unity. Given the limited number of points, the shape of iso-segregation curves cannot be determined. Points with equivalent segregation coefficient are thus joined with dashed lines. The grey zone in the upper part of the diagram indicates a (concentration-stress) zone where no segregation of B is expected. Considering the former comments on iso-segregation curves, its border has a very speculative character.

**Figure 5.** Depth concentration profiles of B in an unstressed $Si_{0.91}Ge_{0.09}$ layer before (full line) and after (triangles) annealing at 900°C during 1h. The simulated profile allowing the measurement of the lattice diffusion coefficient is given for comparison (open circles).

**Figure 6.** Lattice diffusion coefficient of B at 900°C versus Ge concentration (x) in $Si_{1-x}Ge_x$ layers under compression (full squares) and unstressed (full circles). The error bar presented corresponds to the maximum error due to SIMS measurements. It takes into account the variations of the height and the width of profiles measured in a same sample (layer inhomogeneity and SIMS reproducibility).

**Figure 7.** Lattice diffusion coefficient of B at 900°C versus biaxial pressure in $Si_{0.91}Ge_{0.09}$ layers. The error bar presented corresponds to the maximum error due to SIMS measurements. It takes into account the variations of the height and the width of profiles measured in a same sample (layer inhomogeneity and SIMS reproducibility).

**TABLE I**

Lattice diffusion coefficient of B (cm$^2$/s) at 900°C in Si$_{1-x}$Ge$_x$ layers with different Ge concentrations and stress states.

|  | Si | Si$_{0.91}$Ge$_{0.09}$ | Si$_{0.82}$Ge$_{0.18}$ |
|---|---|---|---|
| Under biaxial compression Epitaxy on Si(100) |  | 4.8x10$^{-16}$ | 3.0x10$^{-16}$ |
| Unstressed | 7.1x10$^{-15}$ | 3.9x10$^{-15}$ | 1.2x10$^{-15}$ |
| Under biaxial tension epitaxy on Si$_{0.82}$Ge$_{0.18}$(100) |  | 6.6x10$^{-15}$ |  |

**FIGURE 1.**

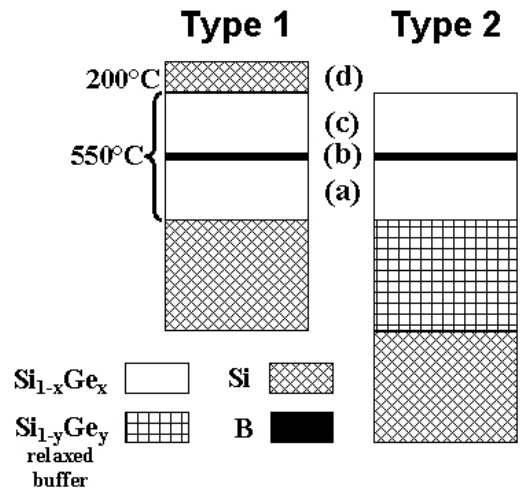

**FIGURE 2.**

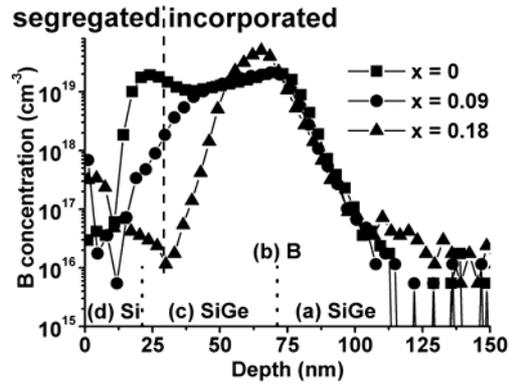

**FIGURE 3.**

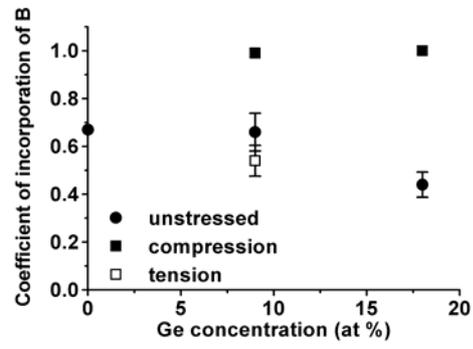

**FIGURE 4.**

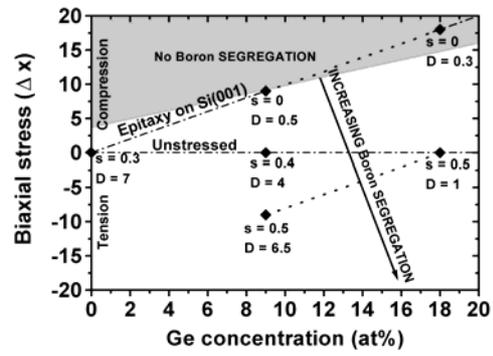

**FIGURE 5.**

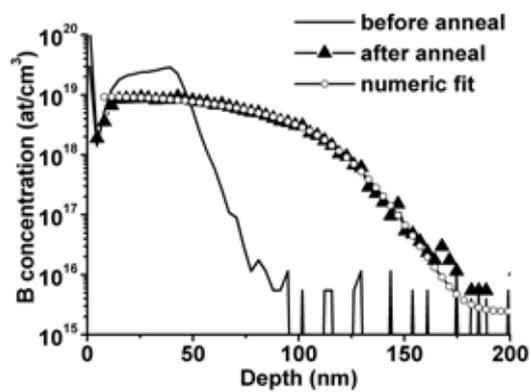

**FIGURE 6.**

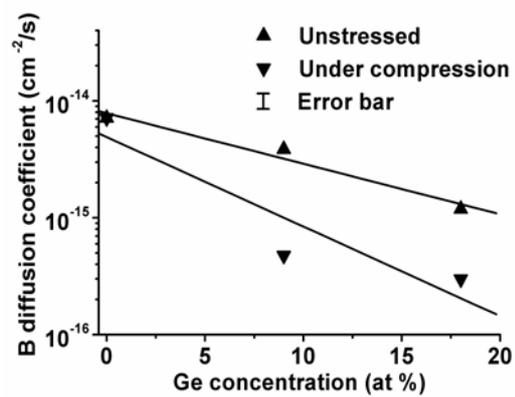

**FIGURE 7.**

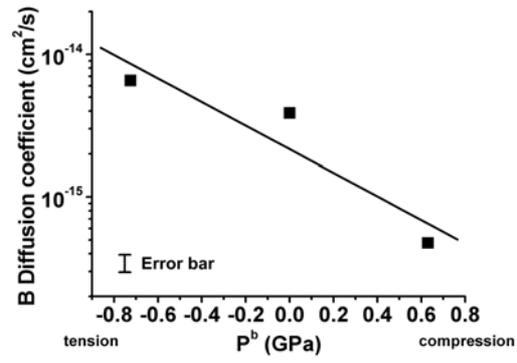